\documentclass{aa}
\usepackage[varg]{txfonts}
\usepackage{natbib}
\bibpunct{(}{)}{;}{a}{}{,} % to follow the A&A style

\usepackage{aas_macros}
\usepackage{amsmath}

\usepackage{color}

\newcommand{\equ}[1]{eq.~(\ref{eq:#1})}

\newcommand{\se}[1]{\S\ref{sec:#1}}
\newcommand{\fig}[1]{Fig.~\ref{fig:#1}}

\newcommand{\Fig}[1]{Figure~\ref{fig:#1}}

\newcommand{\be}{\begin{equation}}
\newcommand{\ee}{\end{equation}}
\newcommand{\ba}{\begin{align}}
\newcommand{\ea}{\end{align}}
\newcommand{\bad}{\begin{equation} \begin{aligned}}
\newcommand{\ead}{\end{aligned} \end{equation}}
\newcommand{\bea}{\begin{eqnarray}}
\newcommand{\eea}{\end{eqnarray}}

\newcommand{\msun}{M_\odot}

\newcommand{\ifm}[1]{\relax\ifmmode#1\else$\mathsurround=0pt #1$\fi}
\newcommand{\kms}{\ifmmode\,{\rm km}\,{\rm s}^{-1}\else km$\,$s$^{-1}$\fi}

\newcommand{\kpc}{\,{\rm kpc}}
\newcommand{\pc}{\,{\rm pc}}
\newcommand{\Gyr}{\,{\rm Gyr}}

\newcommand{\yr}{\,{\rm yr}}
\newcommand{\ergs}{\,{\rm erg}\,{\rm s}^{-1}}

\newcommand{\ltsima}{$\; \buildrel < \over \sim \;$}
\newcommand{\lsim}{\lower.5ex\hbox{\ltsima}}
\newcommand{\gtsima}{$\; \buildrel > \over \sim \;$}
\newcommand{\gsim}{\lower.5ex\hbox{\gtsima}}
\newcommand{\prop}{\propto}

\def\Mv{M_{\rm v}}

\def\Rv{R_{\rm v}}
\def\Vv{V_{\rm v}}

\def\Mg{M_{\rm g}}
\def\Ms{M_{\rm s}}

\def\Sig1{\Sigma_1}

\def\Vsn{V_{\rm SN}}

\def\Mdot{\dot{M}}

\def\tdep{t_{\rm dep}}

\def\tinf{t_{\rm inf}}

\def\Mbh{M_{\rm bh}}
\def\Mcrit{M_{\rm crit}}

\begin{document}

%\large

\title{Origin of the Golden Mass of Galaxies and Black Holes}
\author{Avishai Dekel\inst{1}\inst{2} 
\and Sharon Lapiner\inst{1} \and Yohan Dubois\inst{3}
}
\institute{Racah Institute of Physics, The Hebrew University,
Jerusalem 91904 Israel\\
\email{dekel@huji.ac.il}
\and
SCIPP, University of California, Santa Cruz, CA 95064, USA
\and 
Institut d'Astrophysique, 98 bis Boulevard Arago, 75014 Paris, France
}
\abstract
{}
{We address the origin of the golden mass and time for galaxy formation and
the onset of rapid black-hole growth.
The preferred dark-halo mass of $\sim\!10^{12}\msun$  
is translated to a characteristic epoch, $z\!\sim\!2$, at which the
typical forming halos have a comparable mass.}
{We put together a coherent picture based on existing and new simple analytic
modeling and cosmological simulations.}
{We describe how the golden mass
arises from two physical mechanisms that suppress gas supply and star formation
below and above the golden mass, supernova feedback and virial shock heating of
the circum-galactic medium (CGM), respectively. 
Cosmological simulations reveal
that these mechanisms are responsible for a similar favored mass for the
dramatic events of gaseous compaction into compact star-forming ``blue
nuggets", caused by mergers, counter-rotating streams or other mechanisms.
This triggers inside-out quenching of star formation, to be maintained by the
hot CGM, leading to today's passive early-type galaxies.
The blue-nugget phase is responsible for transitions in the galaxy structural, 
kinematic and compositional properties, e.g., from dark-matter to baryon 
central dominance and from prolate to oblate shape.  
The growth of the central black hole is suppressed by supernova feedback 
below the critical mass, and is free to grow once the halo is massive
enough to lock the supernova ejecta by its deep potential well and the hot CGM.
A compaction near the golden mass makes the black hole sink to the galactic
center and triggers a rapid black-hole growth.
%A compaction near the golden mass is the trigger for rapid black-hole growth 
%through its sinkage to the galactic center. 
This ignites feedback by the Active Galactic Nucleus that helps keeping the 
CGM hot and maintaining long-term quenching.}
{}


\titlerunning{Golden mass}
\authorrunning{Dekel, Lapiner, Dubois}

\maketitle

%%%%%%%%%%%%%%%%%%%%%%% 1
\section{Introduction}
\label{sec:intro}

\begin{figure}
\vskip 5.8cm
\includegraphics{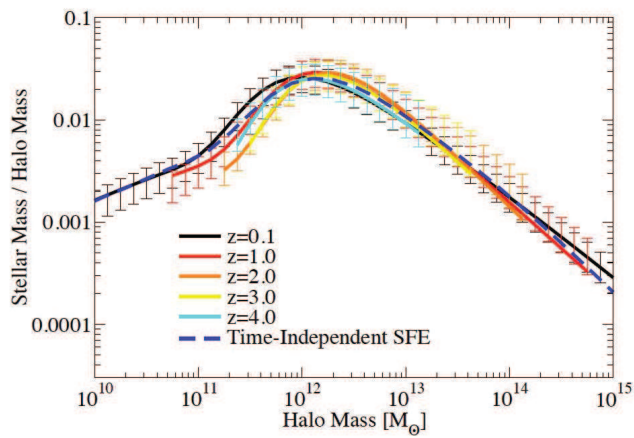}
\caption{
The efficiency of galaxy formation, represented by the ratio of stellar-to-halo
mass, versus halo mass, as derived from
observations by abundance matching to LCDM dark-matter halos
\citep{behroozi13}.
Quenching at low masses and at high masses define a peak near a golden mass
of $\Mv\!\sim\!10^{12}\msun$, roughly the same at all redshifts in the range
$z\!=\!0\!-\!4$ (and at higher redshifts).
One should mention that AGN are preferentially detected above the golden mass
\citep[][see the right panel of \fig{Mdotbh} below]{forster18b}.
}
\label{fig:behroozi}
\end{figure}

% MsMv
The stellar-to-halo mass ratio in galaxies of a given dark-matter halo mass,
$\Ms/\Mv$, can be interpreted as the efficiency of galaxy formation in such 
halos.  This ratio has been derived by abundance matching of observed galaxies 
and simulated LCDM halos \citep{moster10,behroozi13,behroozi18}.
\Fig{behroozi} reveals a robust shape as a function of halo mass, 
with only little redshift dependence in the redshift range $z\!=\!0\!-\!4$.
It shows a peak of efficiency near a halo virial mass 
$\Mv\!\sim\!10^{12}\msun$,
declining on both sides toward lower and higher masses,
indicating the operation of mechanisms that suppress star formation
in the two zones below and above the critical mass.

\begin{figure}
\vskip 6.5cm
\includegraphics{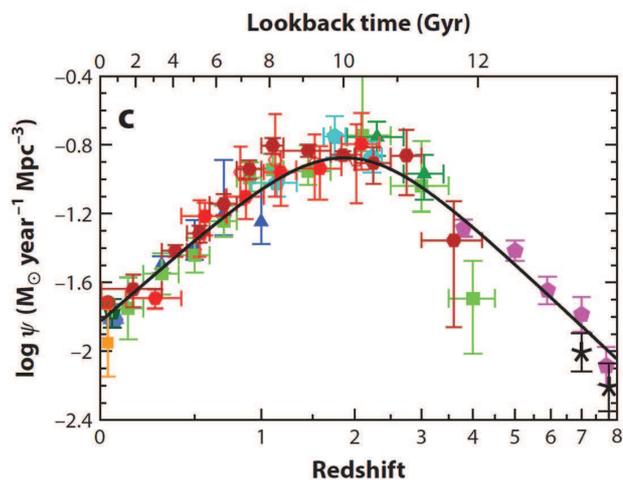}
\caption{
The star-formation-rate density $\psi$ as a function of redshift 
\citep{madau14}, showing
a characteristic epoch for galaxy formation, near $z\!\sim\!2$.
This peak reflects the golden mass for galaxy-formation efficiency from
\fig{behroozi}, convolved with the fact that the typical mass for forming
halos, as evaluated by the Press-Schechter formalism,
is comparable to the
golden mass at $z\!\sim\!2$. The peak in time is thus a result of low-mass
quenching at high redshifts, and of the natural decline of specific accretion
rate with time aided by high-mass quenching at low redshifts.
}
\label{fig:madau}
\end{figure}

\smallskip % Madau and PS
A complementary fundamental observational input is the cosmological evolution
of star-formation rate (SFR) density \citep{madau14}, shown in \fig{madau}.
It rises continuously from
$z\!\sim\!10$ to $z\!\sim\!2$, where it reaches a peak, followed by a steep 
decline after $z\!\sim\!1$ till the present.
We argue that this critical time, at $z\!\sim\!2\!-\!1$,
reflects the same critical mass scale that is
indicated by the stellar-to-halo ratio.
This is because, as obtained from the Press-Schechter formalism 
\citep[][PS]{press74} and confirmed in cosmological simulations,
the typical halos that form at $z\!\sim\!1\!-\!2$ are of $10^{11-12}\msun$, 
and this typical halo mass is very steeply rising with time.

\smallskip
The density of accretion-rate into halos is declining with time at
all epochs, reflecting the expansion of the Universe and the evolution of
the halo population.
This can be derived from the average specific accretion rate,
which in the Einstein-deSitter regime
(at $z\!>\!1$) is $\Mdot/M \!\prop\! (1+z)^{5/2}$ \citep{dekel13},
convolved with the number density of halos based on the Press-Schechter
formalism.
The quoted accretion rate emerges in a simple way from its invariance
with respect to the ``time" variable $(1+z)$, the inverse of the growing mode
of linear fluctuations.

\smallskip
The rise in time of the SFR density prior to the peak at $z\!\sim\!2$, despite
the decline in accretion rate, is obtained by the suppression
of galaxy formation in low-mass halos that are of the Press-Schechter mass
at those high redshifts \citep{bouche10}.
The drop in SFR density with time at low redshifts,
while it qualitatively follows the basic decline in the accretion rate,
is significantly steeper, indicating high mass quenching when the
Press-Schechter mass is high.
Thus, the rise and fall of the SFR density with time is largely determined
by the same quenching mechanisms that operate at low and high mass scales
and generate the peak in $\Ms/\Mv$.
The peak epoch of SFR density can thus be interpreted as a manifestation
of the critical mass scale for peak efficiency of galaxy formation, which we
address here.

\smallskip % bimodality DB06
The golden mass marks a general bimodality about $\Ms\!\sim\!10^{10.5}\msun$
in many galaxy properties, as summarized in \citet[][hereafter DB06]{db06}.
This includes the characteristics of star formation, morphology, kinematics
and composition.
DB06 highlighted the key role played by the virial shock heating of the
Circum-Galactic Medium (CGM), valid in halos above the critical mass,
in generating the bimodality.
This will be summarized below (\se{cgm}) as one of the fundamental
process responsible for the golden scale, to be integrated with another basic
process, supernova (SN) feedback, 
which dominates below the critical mass (\se{sn}).

\smallskip %AGN
Also pointing to the same golden mass is the fact that
Active Galactic Nuclei (AGN) are observed to dominate the
emission lines in galaxies above a threshold mass, roughly
$\Ms \geq 10^{10}\msun$, corresponding to $\Mv \geq 10^{12}\msun$
\citep{kauffmann03_agn,vitale13}.
A similar threshold mass is observed for AGN-driven outflows at high redshifts
\citep{forster18b}.
There are indeed preliminary indications that the masses of black holes
below $10^6\msun$ fall short of the standard linear relation between black-hole
 mass and its host bulge mass \citep{kormendy13,reines15}.
These indicate that black-hole growth is suppressed in the low-mass zone
and it becomes rapid in the high-mass zone,
which also implies that AGN feedback is not relevant in low-mass galaxies
while it may contribute to the quenching of high-mass galaxies.
There seems to be no feature of black-hole physics that may hint to the
origin of such a mass threshold in the black-hole abundance, so one may
suspect that it arises from the cross-talk between the black hole and other
physical processes in the host galaxy that control the gas supply to the black
hole.
Understanding the origin of this interplay between galaxy and black-hole
evolution is a major goal of galaxy formation, which we
attempt to address here.

\smallskip %Compaction
Both simulations \citep{zolotov15,tacchella16_prof,tacchella16_ms}
and observations \citep{barro13,dokkum15,barro17}
reveal a drastic sequence of events that
typically occur in galaxies when they are near the same critical mass
\citep{tomassetti16,huertas18}.
The galaxies undergo dramatic gaseous {\it compactions}
into compact star-forming ``blue nuggets".
The major compaction events trigger inside-out quenching of star formation,
which is maintained by the hot CGM, leading to today's passive elliptical
galaxies.
The blue-nugget phase is responsible for drastic transitions in the main galaxy
structural, kinematic and compositional properties 
(Dekel et al. 2019, in prep.), 
for example a transition from dark-matter to 
baryon dominance in the galaxy center.

\smallskip % What we do here
Here we address the interplay between supernova feedback, CGM heating and
wet compaction in generating the golden scale for galaxy formation and
activating rapid black-hole growth.
We propose that
the supernova feedback and the hot CGM define the zones of low-mass and
high-mass quenching, as well as the critical mass for efficient compaction
between the supernova zone and the hot-CGM zone.
In turn, the compaction event causes the transitions in galaxy properties
at this mass scale. In particular, it boosts the black hole growth, which then,
through AGN feedback, helps the hot CGM maintain the quenching of star
formation above this critical mass.

\smallskip % outline
In \se{sn} we revisit the scale arising from supernova feedback,
and in \se{cgm} the scale associated with a hot CGM.
In \se{compaction} we summarize the event of wet compaction and its
  characteristic scale.
In \se{quenching} we address the quenching mechanism above the critical mass.
In \se{bh} we show how the interplay between the above processes and
black-hole growth imprints the golden mass in the black-hole and AGN
population.
In \se{conc} we conclude our findings.

%%%%%%%%%%%%%%%%%%%%%%%%%% 2
\section{The Supernova-Feedback Scale}
\label{sec:sn}

% \adr{Consider more details from DW03} 

% characteristic SN scale
An upper limit for the dark-matter halo mass (actually its virial velocity)
within which supernova feedback can be effective in heating or ejecting the gas
and thus suppressing the SFR can be estimated in a simple but robust way 
using the standard theory for supernova bubbles \citep[e.g.][]{ds86}.
The energy deposited in the ISM by supernovae that arise from a stellar mass
$\Ms$ is estimated to be 
\be
E_{\rm SN} \sim \Ms \Vsn^2 \, , \quad
\Vsn \sim 120 \kms \, .
\label{eq:Esn}
\ee
The derivation involves the ratio of two timescales, a cooling time and a
dynamical time, which turns out to be roughly constant. These are 
the duration of the adiabatic phase of the supernova bubble 
in which it can deposit energy in the ISM before it cools radiatively,
and the timescale associated with star formation that is roughly proportional
to the dynamical timescale in the star-forming region.

\smallskip
For the supernova energy to heat or eject most of the gas of mass $\Mg$
that has accreted into the galaxy it should be comparable to the binding
energy of this gas in the dark-matter halo potential well,
\be
E_{\rm CGM} \sim \Mg \Vv^2 \, , \quad
\Vv^2 = \frac{G\Mv}{\Rv} \, , 
\label{eq:Ecgm}
\ee  
where $\Mv$ and $\Rv$ are the halo virial mass and radius,
$\Vv$ is the halo virial velocity, and $\Vv^2$ characterizes the halo potential
well. 
At the peak of star-formation efficiency, if a large fraction of the gas
accreted into the halo turned into stars with little ejection, one has 
$\Ms\!\sim\!\Mg$. 
By comparing \equ{Esn} to \equ{Ecgm},
this yields a critical upper limit for the virial velocity of a halo
in which supernova feedback can be effective,
\be
\Vv \sim \Vsn \sim 120 \kms \, .
\ee 
Using the standard virial relation,
this corresponds to $\Mv\!\sim\!10^{11.7}\msun$ at $z\!=\!0$,
and it roughly scales as $(1+z)^{-3/2}$ in the EdS regime
\citep[e.g.][Appendix A2]{db06}. 
In haloes above this mass, the potential well is too deep for the
supernova-driven winds to escape or for a large fraction of the gas to heat up.
This mass scale roughly coincides with the observed peak of star-formation
efficiency.

\begin{figure}
\vskip 7.55cm
\includegraphics{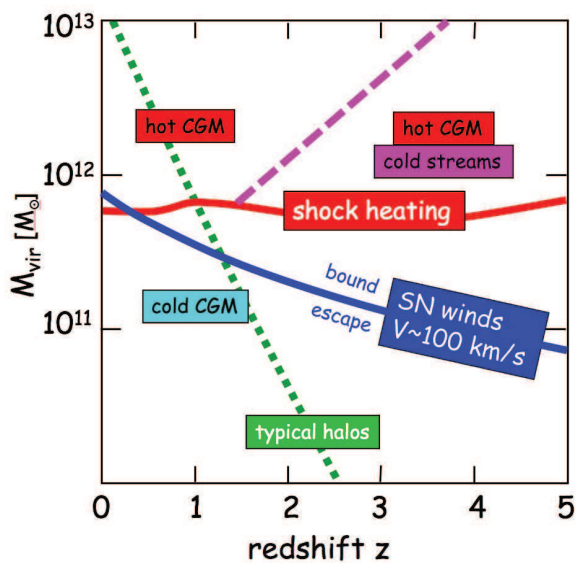}
\caption{
The characteristic halo masses as a function of redshift.
The blue curve marks the upper limit for effective supernova feedback
based on \citet{ds86}(\se{sn}).
The red curve denotes the threshold for virial shock heating of the CGM
based on \citet{db06}(\se{cgm}),
with the dashed magenta line representing the upper limit for
penetrating cold streams.
The green short-dash curve refers to the Press-Schechter mass,
the typical mass of forming halos.
The left portion of the red curve at $z\!<\!2$, combined with the dashed
magenta curve at $z\!>\!2$, mark the upper limit for star-forming galaxies.
The three masses roughly coincide at $z\!\sim\!1\!-\!2$, defining the peak
epoch and mass of star formation.
}
\label{fig:scale_z}
\end{figure}

\smallskip %fig
\Fig{scale_z} illustrates the critical halo mass corresponding to 
$\Vv\!=\!\Vsn$ as a function of redshift.
Gradually rising with time,
it coincides at $z\!\sim\!1\!-\!2$ with the steeply-rising Press-Schechter mass
that characterizes the typical mass for haloes at a given time.

\smallskip % slope in the SN zone
This simple energetics of supernova feedback can also be used to predict the
slope of the stellar-to-halo mass relation in the supernova zone below the 
critical mass. When $\Vv \!<\! \Vsn$, the shallow potential well allows 
significant gas ejection, namely $\Ms \!\ll\! \Mg$. 
The gas mass that has been accreted into the halo is assumed to be
roughly proportional to the halo mass, $\Mg \!\prop\! \Mv$. 
Comparing \equ{Esn} and \equ{Ecgm} then yields \citep{dw03}
\be
\frac{\Ms}{\Mv} \prop \Vv^2 \prop \Mv^{2/3} \, .
\ee
This indeed resembles the slope deduced from observations via abundance
matching in the supernova zone (\fig{behroozi}).
Note that in the vicinity of the critical potential well, $\Vv\!\sim\!\Vsn$,
the assumption of no ejection automatically implies a flat peak,
$\Ms/\Mv\!\sim\! const$.

\smallskip
Similar arguments \citep{dw03} lead to the scaling relations of surface
brightness with mass in the supernova, dwarf zone, $\mu \!\prop\! \Ms^{3/5}$,
as well as the metallicity with mass, $Z \!\prop\! \Ms^{2/5}$,
and circular velocity with mass, $V \!\prop\! \Ms^{1/5}$.
These simple arguments predict the flattening off of the surface brightness and
metallicity relations near the critical mass, and the convergence to the
standard Tully-Fisher relation of $V \!\prop\! \Ms^{1/4}$ there.

\smallskip
Supernova feedback indeed plays a major role in galaxy evolution in the
supernova zone, below the golden mass.
For example,
it can explain the formation of flat cores in dark-matter halos
by repeating episodes of star formation and supernovae
\citep{ds86, dw03, pontzen12, freundlich19},
the low surface brightness of dwarf galaxies
including the recently discovered ultra-diffuse
galaxies \citep{dutton16b,dicintio17,jiang19_udg},
the missing dwarfs in the local Group \citep{zolotov12,garrison13},
and the Kennicutt-Schmidt relation between the global
SFR density and gas density \citep{silk97,ostriker11,dekel19}.

%%%%%%%%%%%%%%%%%%%%%%%%%%%%%%%%%% 3
\section{The Hot-CGM Scale}
\label{sec:cgm}

\begin{figure}
\vskip 7.1cm
\includegraphics{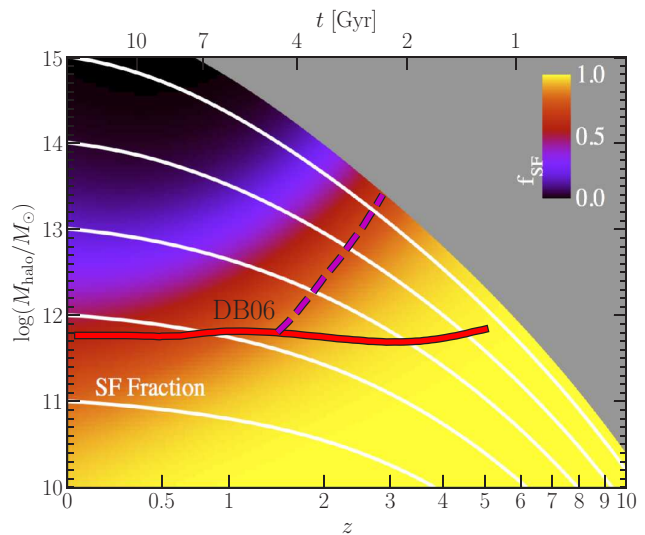}
\caption{
The fraction of star-forming galaxies as a function of halo mass and redshift,
as derived from observed galaxies using abundance matching to LCDM halos
\citep[based on][]{behroozi16}.
The transition from dominance by star-forming versus quenched galaxies (dark
red area) matches the theoretical predictions (red and magenta curves),
the maximum mass indicated by the curves in \fig{scale_z}
\citep[based on][]{db06}.
}
\label{fig:behroozi_17}
\end{figure}

The other characteristic scale is the upper limit for the halo mass within
which efficient cold inflow can supply gas for star formation
\citep{ro77,silk77,binney77}.
It is obtained by comparing the gas radiative cooling time to the relevant
dynamical time for gas inflow.
The key question is whether the shock that forms at the halo virial radius,
behind which the gas heats to the virial temperature,
can be supported against gravitational collapse.
For the post-shock gas to be able to sustain the pressure that supports
the shock against gravity, its cooling time has to be longer than the
dynamical time for gas compression behind the shock, which is comparable to the
halo crossing time, $\Rv/\Vv$ \citep{bd03,db06}.
Since the cooling time is an increasing function of halo mass,
a shock-stability analysis reveals a threshold mass for a hot CGM
on the order of
\be
\Mv \sim 10^{11.7}\msun \, ,
\ee
roughly independent of redshift in the range $z\!=\!0\!-\!3$,
with an uncertainty of a factor of a few due to the uncertainty in metallicity
and the location within the halo where the shock stability is evaluated.
This analysis has been supported by idealized spherical simulations.

%Fig. 7 of DB06
\smallskip
\Fig{scale_z} shows the predicted critical halo mass for virial shock
heating as a function of redshift, based on \citet[][Fig. 7]{db06}.
Below the critical curve one expects the cosmological inflow to be all cold, 
at $T\!\sim\!10^4$K, efficiently feeding the galaxy and allowing high SFR.
Above the curve one expects the CGM to be shock heated to the virial 
temperature, thus suppressing the cold gas supply into the galaxy, and
maintaining long-term quenching.

\smallskip
At $z \geq 2$, above the shock-heating curve and below the dashed curve,
narrow cold streams are expected to
penetrate through the otherwise hot CGM and supply gas for efficient star
formation even in haloes above the critical mass for shock heating.
These predictions, based on an analytic study of virial shock stability and
supporting idealized spherical simulations, have been confirmed in cosmological
simulations \citep[][Fig. xxx]{keres05,ocvirk08,nelson13,nelson16}.

\smallskip % Behroozi 17
These predictions, as summarized in \fig{scale_z}, have been confirmed
observationally.
\Fig{behroozi_17} shows the fraction of star-forming galaxies among the general
population of galaxies in the plane of halo mass versus redshift, similar to
the plane in \fig{scale_z}. 
This is based on abundance matching of galaxies to dark-matter haloes
in a LCDM cosmology \citep{behroozi18}.
The transition from low to high SF fraction, marked
by the red color, matches the predicted critical curves marking the upper
limits for cold gas supply in \fig{scale_z}. 

\smallskip
For long-term quenching, the CGM has to be kept hot. This
can be caused by gravitational heating due to accreting mass \citep{db08},
or by AGN feedback \citep[e.g.][]{croton06,cattaneo07,dubois11},
see \se{quenching} and \se{bh}.

%%%%%%%%%%%%%%%%%%%%%%%%%%%%%%%%%%% 4
\section{Wet Compaction to a Blue Nugget}
\label{sec:compaction}

\begin{figure}
\vskip 6.7cm
\includegraphics{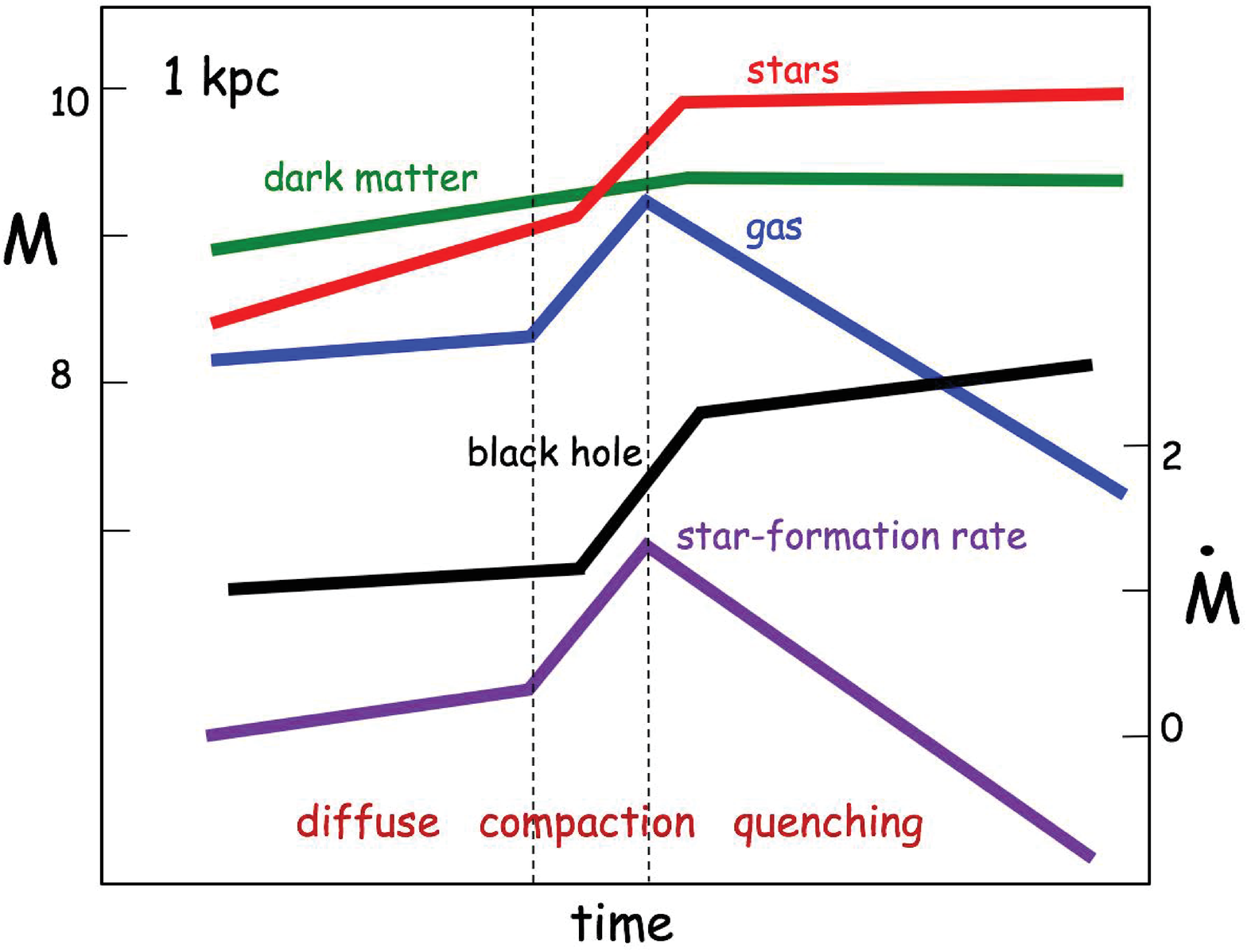}
\caption{
A cartoon describing a typical wet-compaction event as seen in cosmological
simulations, showing the evolution of masses (in $\log (M/\msun)$) 
within the inner $1\kpc$ \citep[based on][]{zolotov15}.
The compaction is the steep rise of gas mass (blue), by an order of magnitude
during $\sim\!0.3\, t_{\rm Hubble}$, reaching a peak as a blue
nugget (BN), and soon after declining as the central gas is depleted by star
formation and outflows with no replenishment.
The SFR (magenta, in $\log (\msun \yr^{-1})$ follows closely,
demonstrating post-BN compaction-triggered central quenching.
The central stellar mass (red) is rising accordingly during the compaction,
and it flattens off post-BN.
The inner $1\kpc$ is dominated by dark matter (green) pre compaction
and by baryons (stars, red) post compaction.
The ``disk" kinematics is dispersion-dominated pre-BN and rotation-dominated
post-BN.
The time of the major blue-nugget event is typically when the galaxy is near 
the golden mass, $\Ms\!\sim\!10^{10}\msun$, separating between the 
pre-compaction supernova phase and the post-compaction hot-CGM phase. 
The black-hole growth (black, \se{bh}), which is suppressed by supernova
feedback pre compaction, is growing during and after the compaction in 
the hot-CGM phase above the golden mass. The onset of rapid black-hole
growth is driven by the compaction event.
}
\label{fig:compaction}
\end{figure}

\begin{figure}
\vskip 7.6cm
\includegraphics{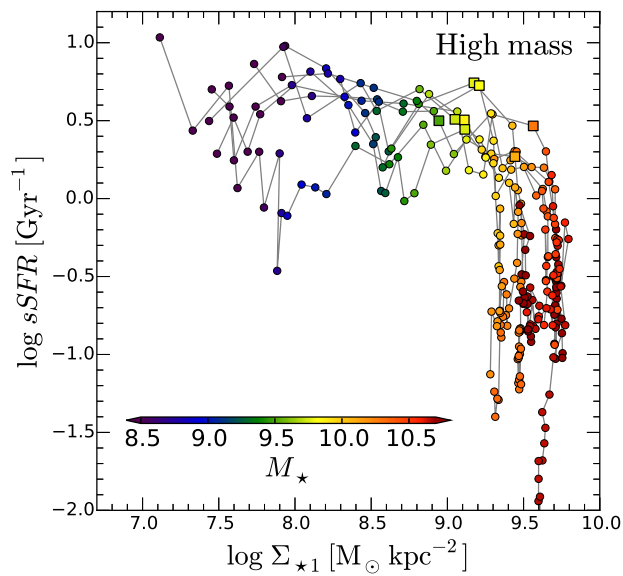}
\caption{
The universal L-shape evolution track of eight galaxies from the VELA 
zoom-in cosmological simulations in the
plane of sSFR and stellar surface density within $1\kpc$, 
$\Sigma_1$, which serves as a measure of compactness 
(Dekel et al. 2019, in prep.). 
The compactness is growing at a roughly constant sSFR (horizontally)
before and during the compaction event, turning over at the blue-nugget phase
(the ``knee", marked by a square symbol) to quenching at a constant 
$\Sigma_1$ (vertically).
A similar behavior is seen observationally \citep[][Fig.~7]{barro17},
with the value of $\Sigma_1$ at the blue-nugget phase weakly increasing with 
redshift.
}
\label{fig:L_shape}
\end{figure}

\smallskip % compaction events
Cosmological simulations show that most galaxies evolve through a dramatic
wet-compaction event, which tends to occur at its maximum strength when the
galaxy mass is near or above the golden value, $\Mv\!\sim\! 10^{12}\msun$ and
$\Ms\!\sim\! 10^{10}\msun$,
especially in the redshift range $z\!=\!1\!-\!5$ when the gas fraction is high
\citep{zolotov15,tacchella16_ms,tomassetti16}.
This event is a significant gaseous contraction into a compact central
star-forming core within the inner $1\kpc$, termed ``blue nugget" (BN).
The gas consumption by star formation and the associated gas ejection by
stellar and supernova feedback lead to central gas depletion and 
inside-out quenching of star-formation rate (SFR) at a roughly constant 
central stellar density \citep{tacchella16_prof}. 
This is illustrated in \fig{compaction}, a cartoon that represents the
evolution seen in many simulated galaxies, showing the evolution of gas mass,
stellar mass and SFR within the inner kiloparsec.
\Fig{L_shape} shows evolution tracks of eight galaxies from the VELA zoom-in
cosmological simulations (with high spatial resolution of $\sim\! 25\pc$ in the
dense regions)
in the plane of specific SFR (sSFR) versus
compactness as measured by the stellar surface density within $1\kpc$,
$\Sigma_1$  
\citep{zolotov15,lapiner19}.
A compaction at a roughly constant sSFR turns into a quenching at a constant
$\Sigma_1$ , generating an L-shape evolution track with the "knee" marking the 
blue-nugget phase.
This characteristic L-shape evolution track has been confirmed observationally
\citep[][Fig.~7]{barro17}. 

\smallskip % obs  
It became evident that the observed massive, passive galaxies, which are
already abundant at $z\!\sim\! 2\!-\!3$, are compact (or have compact cores), 
encompassing $\sim\! 10^{10}\msun$ of stars within $1\kpc$, 
termed ``red nuggets" 
\citep{dokkum08,damjanov09,newman10,
dokkum10,damjanov11,whitaker12,bruce12,dokkum14,dokkum15}.
The effective radii of these compact stellar systems, with respect to their
halo virial radii (indicated, e.g., from the universal stellar-to-halo mass
relation), are typically of order $\sim\!0.01$. 
This is smaller than one would expect had the gas started in the halo with a 
standard spin of $\lambda \!\sim\! 0.035$ and conserved angular momentum during 
the infall into the central galaxy,
thus indicating dissipative inflow associated with
angular-momentum loss -- a wet compaction \citep{db14}.
This implies the presence of blue nuggets as the immediate progenitors of the 
red nuggets.
Indeed, star-forming blue nuggets have been convincingly observed, with 
masses, structure, kinematics and abundance consistent with being the 
progenitors of the red nuggets
\citep{barro13,barro14_bn_rn,barro14_kin,williams14,barro15_kin,dokkum15,
williams15,barro16_alma,barro16_kin,barro17_alma,barro17}.
In particular,
a machine-learning study, after being trained on the blue nuggets as 
identified in the simulations, recognized with high confidence similar blue 
nuggets in the CANDELS-HST multi-color imaging survey of $z\!=\!1\!-\!3$ 
galaxies \citep{huertas18}.

\smallskip % trigger
The compaction process requires a significant dissipative angular-momentum
loss. This is found in the simulations to be caused either by wet mergers
($\sim\! 40\%$ by major plus minor mergers), 
by colliding counter-rotating streams, by recycling fountains
and by other processes (Dekel et al. 2019, in prep.),
and to be possibly associated with violent disk instability \citep{db14}.
These processes preferentially occur at high redshifts,
where the overall accretion is at a higher rate and more gaseous,
leading to deeper compaction events.

\smallskip % transitions
The compaction to blue-nugget event marks drastic transitions in the galaxy
structural, compositional, kinematic and other physical properties, which
translate to pronounced changes as a function of mass near the critical mass
\citep{zolotov15,tacchella16_prof,tacchella16_ms}.
As mentioned, the compaction triggers inside-out quenching of star formation,
to be maintained by a hot CGM in massive halos possibly aided by AGN feedback
(\se{quenching}).
This is accompanied by a structural transition from a diffuse and largely
amorphous configuration to a compact system, possibly surrounded by an 
extended gas-rich ring and/or a stellar envelope.
The kinematics evolves accordingly from pressure to rotation support.
Most noticeable is a compaction-driven transition from central dark-matter
dominance to baryon dominance, which induces a transition of global
shape from a prolate to an oblate stellar system \citep{tomassetti16}.
Finally, the blue nugget marks a transition in the central black-hole growth
rate from slow to fast (\se{bh}), which induces a transition from supernova
feedback to AGN feedback as the main source for quenching.

\smallskip % compaction at a critical mass
According to the simulations, minor compaction events may occur at all masses
in the history of a star-forming galaxy (SFG).
Indeed, repeated episodes of compactions and subsequent quenching attempts
can explain the confinement of SFGs to a narrow Main Sequence 
\citep{tacchella16_ms}.
However, the major compaction events, those that involve an order-of-magnitude
increase in central density, cause a transition from central dark-matter
dominance to baryon dominance, and trigger significant quenching, 
are predicted by the simulations to occur near the golden mass, 
\citet[][Fig.~8]{tomassetti16} and \citet[][Fig.~21]{zolotov15}.
This has been confirmed by the deep-learning study of VELA simulations
versus observed CANDELS galaxies \citep{huertas18}, which
detected a preferred stellar mass for the observed blue nuggets
near the golden mass, $\Ms \!\sim\! 10^{10}\msun$. 
The significance of this finding is strengthened by the fact that
the same characteristic mass has been recovered after eliminating from the
training set the direct information concerning the mass, through the galaxy 
luminosity. 

\smallskip % origin of critical mass by SN
We argue that the origin of the favored mass for major compaction events
is rooted in the two basic processes addresses in \se{sn} and \se{cgm}, namely
the supernova feedback that dominates below the critical mass and
the hot CGM that dominates above the critical mass.
Central supernova feedback, that is boosted following the SFR
soon after the early phases of the compaction process,
removes central gas and halts further compaction in
galaxies below the critical mass for efficient supernova feedback.
Supporting evidence for the effect of supernova feedback on the depth of
compaction events is provided by simulations with stronger feedback, which 
indeed show compaction events that are less dramatic 
(NIHAO, New Horizon, in prep.).
Near and above the critical mass, where the potential well is deep enough and
becomes even deeper due to the compaction itself, the compaction is not
significantly affected by supernova feedback and it can proceed to higher
central densities.
Thus, we propose that
major compactions tend to occur in galaxies near the critical mass
primarily due to the fact that supernova feedback becomes inefficient near 
and above this mass.

%%%%%%%%%%%%%%%%%%%%%%%%%%% 5
\section{Quenching Trigger \& Maintenance}
\label{sec:quenching}

\begin{figure}
\vskip 6.3cm
\includegraphics{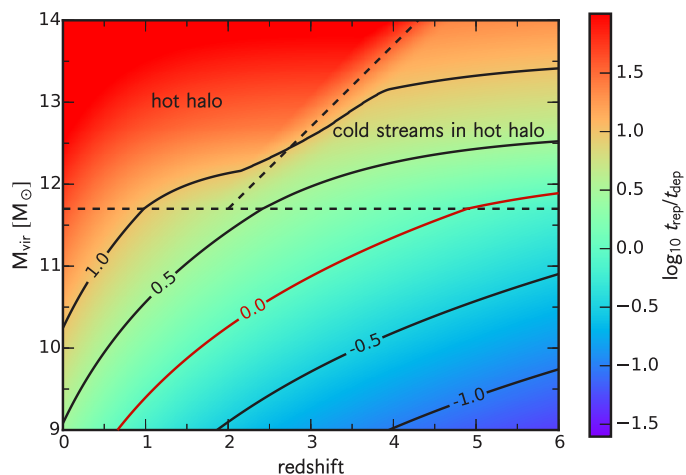}
\caption{
The ratio of timescales for gas inflow-replenishment and for depletion from the
inner $1\kpc$, in the halo mass versus redshift plane,
distinguishing between necessary conditions for wet compactions
and star formation when $\tinf/\tdep \!<\!1$ and deep quenching when
$\tinf/\tdep \!>\! 1$ \citep[following][]{tacchella16_ms}.
Resembling the characteristic masses deduced from \fig{scale_z},
the border line $\tinf/\tdep \!=\!1$ is near the golden mass at $z \!\leq\! 2$,
rising toward higher redshifts to allow for cold streams,
consistent with \fig{behroozi_17} derived from observations.
}
\label{fig:tacchella}
\end{figure}

In order to understand the evolution within the central $1\kpc$ of a galaxy,
and in particular the quenching process at and above the critical mass,
it is helpful to appeal to the key parameter $\tinf/\tdep$,
the ratio of timescales for gas inflow into this region and
gas depletion from it, by star formation and outflows \citep{tacchella16_ms}.
Wet compaction is possible if $\tinf\!<\!\tdep$, such that significant 
dissipative compaction can occur before the gas would turn to stars or be 
ejected by feedback \citep{db14}.
This is also the condition for halting an ongoing depletion-quenching event and
starting a new compaction event by newly accreted gas.
On the other hand, the condition for quenching is $\tdep\!<\!\tinf$, ensuring 
that significant depletion occurs before the gas may be replenished by 
accretion.

\smallskip
In the cold-flow regime below the critical mass for virial shock heating
(\se{cgm}),
the specific inflow rate into the galaxy roughly follows that of the
cosmological total specific accretion rate into the halo.
In the EdS regime (roughly $z\!>\!1$) this is derived analytically and 
confirmed by simulations \citep{dekel13} to be
\be
\tinf \sim 25 \Gyr\, (1+z)^{-5/2}\, M_{12}^{0.14} \, ,
\ee
where $M_{12}$ is the halo mass in $10^{12}\msun$.
It implies that below the critical mass $\tinf$ is a strong function of
redshift and a weak function of mass.
The inflow time becomes much longer once the halo is above the critical
mass, where the heated CGM suppresses the cold gas supply,
and especially at $z \!<\! 2$, when the penetration of cold streams through the
hot CGM is suppressed \citep{db06,cattaneo06}.

\smallskip
The average depletion time near the ridge of the Main Sequence of SFGs,
as estimated from simulations \citep{tacchella16_ms} and observations
at $z\!=\!1\!-\!3$ \citep{tacconi18}, is approximately
\be
\tdep \sim 1 \Gyr\, (1+z)^{-0.5} \, M_{12}^{-0.2}\, .
\ee
Here the depletion time is a weak function of both redshift and mass.
We thus have below the critical mass
\be
\frac{\tinf}{\tdep} \sim 25\, (1+z)^{-2}\, M_{12}^{0.34} \, .
\ee
For galaxies of $M_{12} \!\sim\! 0.3$, say, we have $\tinf \!\sim\! \tdep$
at $z \!\sim\! 3$.
At higher redshifts it is more likely to have $\tinf/\tdep \!<\!1$,
a necessary condition for wet compaction when there is a trigger,
followed by a burst of star formation.
At lower redshifts, where $\tinf/\tdep\!>\!1$ is more common, there is 
efficient post-compaction central gas depletion with no efficient replenishment
by fresh cold gas, allowing deeper long-term quenching, especially when the 
CGM is hot above the critical mass.

\begin{figure*}
\vskip 7.4cm
\includegraphics{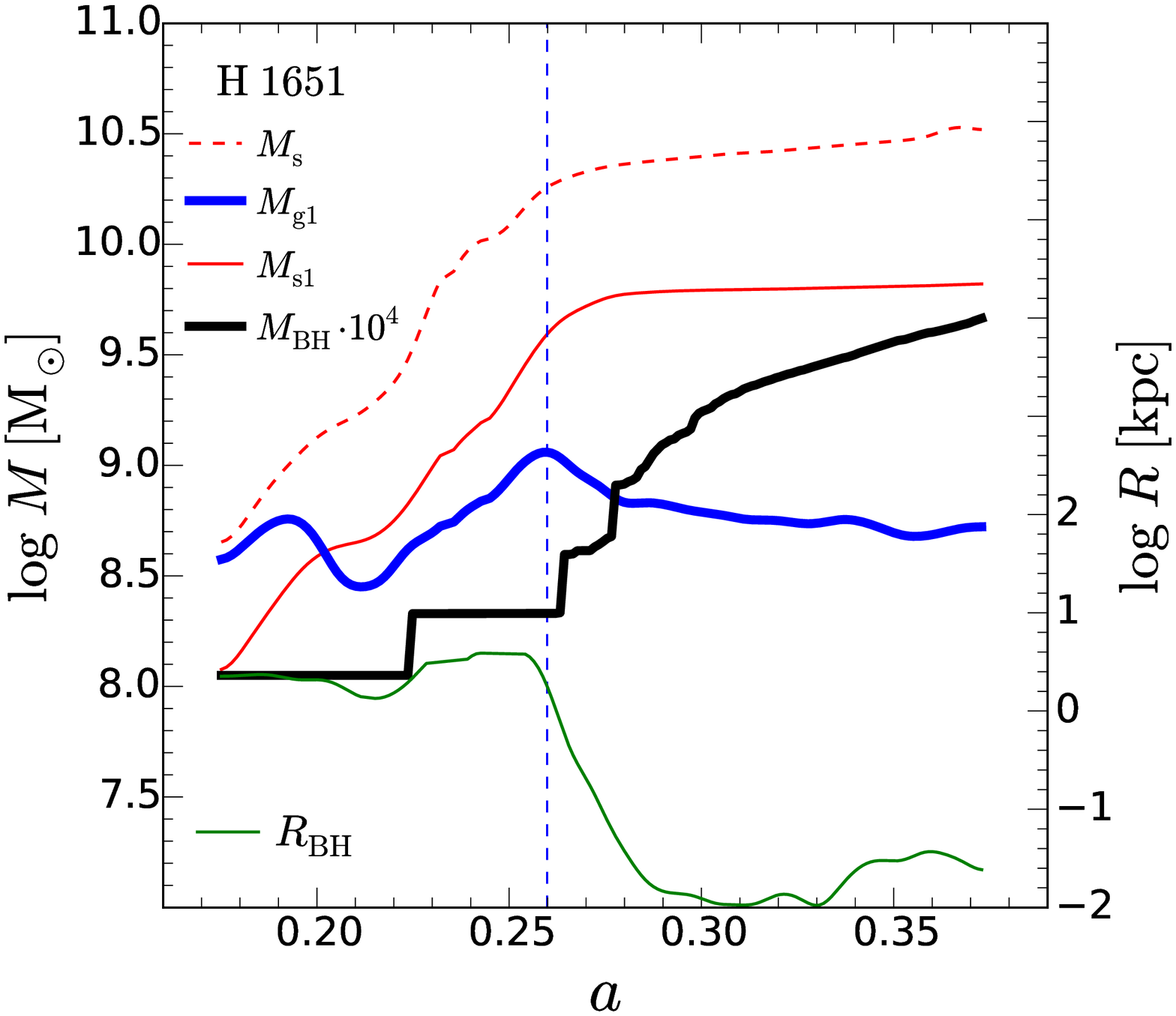}
\includegraphics{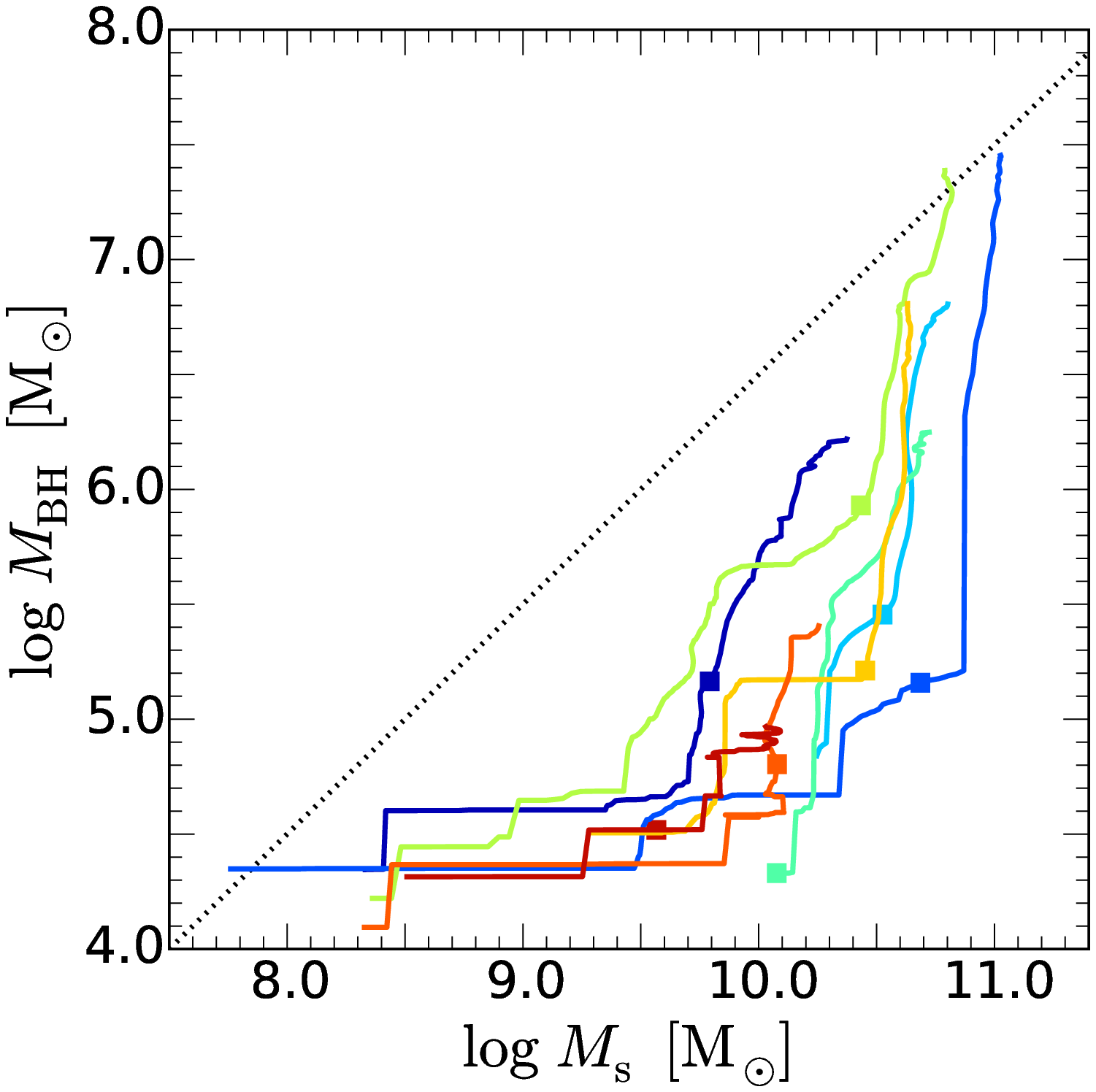}
\caption{
Compaction-driven black-hole growth in New-Horizon simulations
\citep{lapiner19}.
{\bf Left:}
Evolution of black-hole mass ($\times 10^4$, black) and gas mass
within $1\kpc$ (blue) as a function of cosmological expansion factor $a$
in a simulated galaxy.
The black-hole growth is suppressed by supernova feedback at early times.
The compaction event, marked by the rise of the gas mass to a peak,
triggers a rapid black-hole growth.
The galaxy stellar mass (dashed red) at that time is near the critical mass
$\sim\! 10^{10.3}\msun$.
The compaction brings the black hole to the galaxy center (green, $\log R$).
{\bf Right:}
Evolution tracks of black-hole mass versus stellar mass $\Ms$ in eight
galaxies.
They show supernova-driven suppression of black-hole growth below
a critical mass of $\Ms\!\sim\!10^{10}\msun$,
turning into a rapid growth near the critical mass,
likely driven by compaction events (squares).
Similar L-shape tracks are seen when $\Mbh$ is plotted against bulge mass or
$\Sigma_1$.
Black holes of $\sim\!10^{5}\msun$ %in galaxies of $\Ms\!\sim\!10^{9.5}\msun$
are thus predicted to lie below the standard linear relation.
}
\label{fig:Mbh}
\end{figure*}

\smallskip
\Fig{tacchella}, borrowed from \citet{tacchella16_ms},
shows the expected quenching efficiency, the ratio $\tinf/\tdep$, as a
function of halo mass and redshift.
There is an encouraging qualitative resemblance between this and
\fig{behroozi_17}, which shows in the same plane the distribution of
star-forming fraction as estimated from observations via abundance matching.

\smallskip
Overall, the quenching mechanism is primarily a function of mass.
In haloes of masses below the critical mass there is quenching by stellar and
supernova feedback. Especially at high redshift, cold gas supply may cause a
new
compaction event and a new burst of star formation, with boosted supernova
feedback
that triggers a new quenching attempt. The associated oscillations of
$\tinf/\tdep$ about unity cause oscillations about the Main Sequence of SFGs,
which can explain the confinement of SFGs to a narrow Main Sequence
\citep{tacchella16_ms}.
Near the critical mass, a major compaction event triggers a central burst of
star formation, which leads to the onset of deeper quenching by gas consumption
and supernova feedback that is not followed by efficient replenishment.
The triggered quenching is maintained in haloes above the critical mass because
the hot CGM suppresses gas supply to the central galaxy. This is especially
efficient at low redshifts ($z\!<\!2$), where cold streams do not bring gas in
above the critical mass \citep{db06}.
As discussed below,
in the hot-CGM regime, a major compaction event also activates a rapid growth
of the central black hole, leading to AGN that helps keeping the
CGM hot and thus maintain the quenching of star formation.

%%%%%%%%%%%%%%%%%%%%%%%%%%%%%%%%%%%%% 6
\section{Black Hole Growth}
\label{sec:bh}

% SN suppression and compaction-driven growth
The evolution of galaxies from the supernova zone below the critical mass,
through a
compaction event near the critical mass, into the hot-CGM phase above the
critical mass, determines the black-hole growth rate and imprints the golden
mass in it.
Below the critical mass, namely pre-compaction, the supernova-driven gas
heating and ejection from the center suppresses the black-hole growth.
The major compaction that tends to occur near the critical mass, where
supernova feedback is already inefficient, brings gas into the inner
sub-kiloparsec blue nugget. This can induce efficient accretion onto the
sub-parsec black hole,
which can trigger rapid black-hole growth and the activation of an AGN.
In more massive haloes the deep potential well and the hot CGM lock the
central gas in, and allow continuing accretion onto the black hole.

\smallskip % sims. New Horizon
The suppression of black-hole growth by supernova feedback below the critical
mass and the transition to rapid black-hole growth above it have been 
demonstrated in several different cosmological simulations,
which were run with and without supernova feedback.
This phenomenon is robust given that the simulations have been using different 
codes and subgrid recipes and especially different implementations of 
supernova feedback and black-hole growth.
It was first seen in a RAMSES simulation by \citet{dubois15} and
\citet{habouzit17},
and then in EAGLE simulations \citep{bower17},
FIRE simulations \citep{angles17} and Illustris TNG \citep{habouzit18}.

\smallskip
This phenomenon, and in particular the crucial role of wet compaction in 
triggering the black-hole growth near the golden mass,
is demonstrated in ten massive galaxies from the New-Horizon cosmological
simulation (Dubois et al. in prep.).
We summarize here the preliminary results, to be analyzed in more detail
in \citep{lapiner19}.

\smallskip % 
\Fig{Mbh} (left panel)
 shows the evolution of black-hole mass (scaled) and the masses of
gas and stars within the inner $1\kpc$ as a function of expansion factor 
$a\!=\!(1+z)^{-1}$
in a New-Horizon simulated galaxy.
The gas (blue) shows a wet-compaction event, with an onset at $a \simeq 0.21$
and a blue-nugget peak at $a \simeq 0.26$.
The history of SFR within the inner $1\kpc$ (not shown) follows closely
the gas history.
The associated stellar mass within $1\kpc$ (red) is rising until the
blue-nugget peak and flattens off thereafter.
The total stellar mass (dashed red) at the blue-nugget peak
is $\simeq 10^{10.15}\msun$, representing the golden stellar mass scale.
We see a suppressed black-hole growth in the pre-compaction supernova regime
and a rapid growth post compaction. The black-hole growth starts at the onset
of compaction, and it continues throughout the compaction process
and later on into the hot-CGM regime.

\smallskip
\Fig{Mbh} (right panel)
puts together eight simulated galaxies, showing the evolution
tracks of black-hole mass versus total stellar mass $\Ms$.
A characteristic L-shape evolution is seen, displaying
a suppression of black-hole growth below the golden mass of
$\Ms \!\sim\! 10^{10}\msun$
and a rapid black-hole growth once above the golden mass.
The blue-nugget phase, marked by a square symbol, typically coincides with the
upturn of the curve, consistent with a causal connection between the two.
This is also demonstrated when $\Mbh$ is plotted against $\Sigma_1$,
the stellar surface density within the inner $1\kpc$,
which serves as a measure of compactness as in \fig{L_shape}.
A similar turnover is found to occur at a characteristic
threshold of $\Sigma_1\!\sim\!10^9 \msun \kpc^{-2}$, consistent with
compaction-driven black-hole growth.
Similar tracks are seen when $\Mbh$ is plotted against the bulge mass,
predicting that black holes of $\sim\! 10^5\msun$ should lie well below
the standard linear relation between black-hole mass and bulge mass that has
been established for black holes more massive than $10^6\msun$
\citep{magorrian98,kormendy13}.

\smallskip % BH radius
A clue for how the compaction triggers the black-hole growth is provided by
tracing the position of the black hole with respect to the galaxy center.
The green curve in \fig{Mbh} shows that during the pre-compaction phase the
black hole orbits at a few kiloparsecs off the center, where the orbital
velocity and the supernova-driven dilute-gas environment suppresses the
accretion onto the black hole. 
Soon after the compaction, the black hole sinks to the
$\sim\!10\pc$ vicinity of the galaxy center, due to drag against the
compaction-driven dense gas and deep potential well once above the golden
mass. Once at rest in the dense galaxy center, the black hole is
subject to efficient accretion.
The wondering of the black holes off the centers has been
also seen in simulations of low-mass galaxies by \citet{bellovary19}.

\begin{figure}
\vskip 6.5cm
\includegraphics{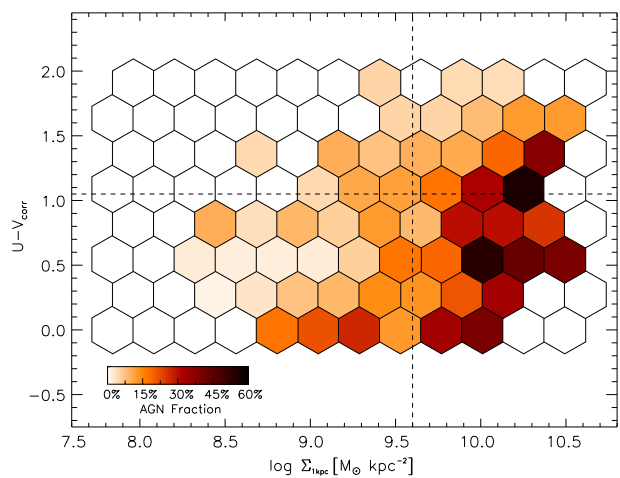}
\caption{
Fraction of CANDELS galaxies that host AGN (color scale)
in the plane of SFR (growing downwards) and compactness,
as represented by U-V color and stellar surface
density inside $1\kpc$ $\Sigma_{\rm 1\kpc}$ respectively
\citep[following][where the effective radius was used]{kocevski17}.
The AGN fraction is high in the blue-nugget quadrant of compact star-forming
galaxies (bottom-right), consistent by compaction-driven compaction.
}
\label{fig:kocevski}
\end{figure}

\begin{figure*}
\vskip 7.4cm
\includegraphics{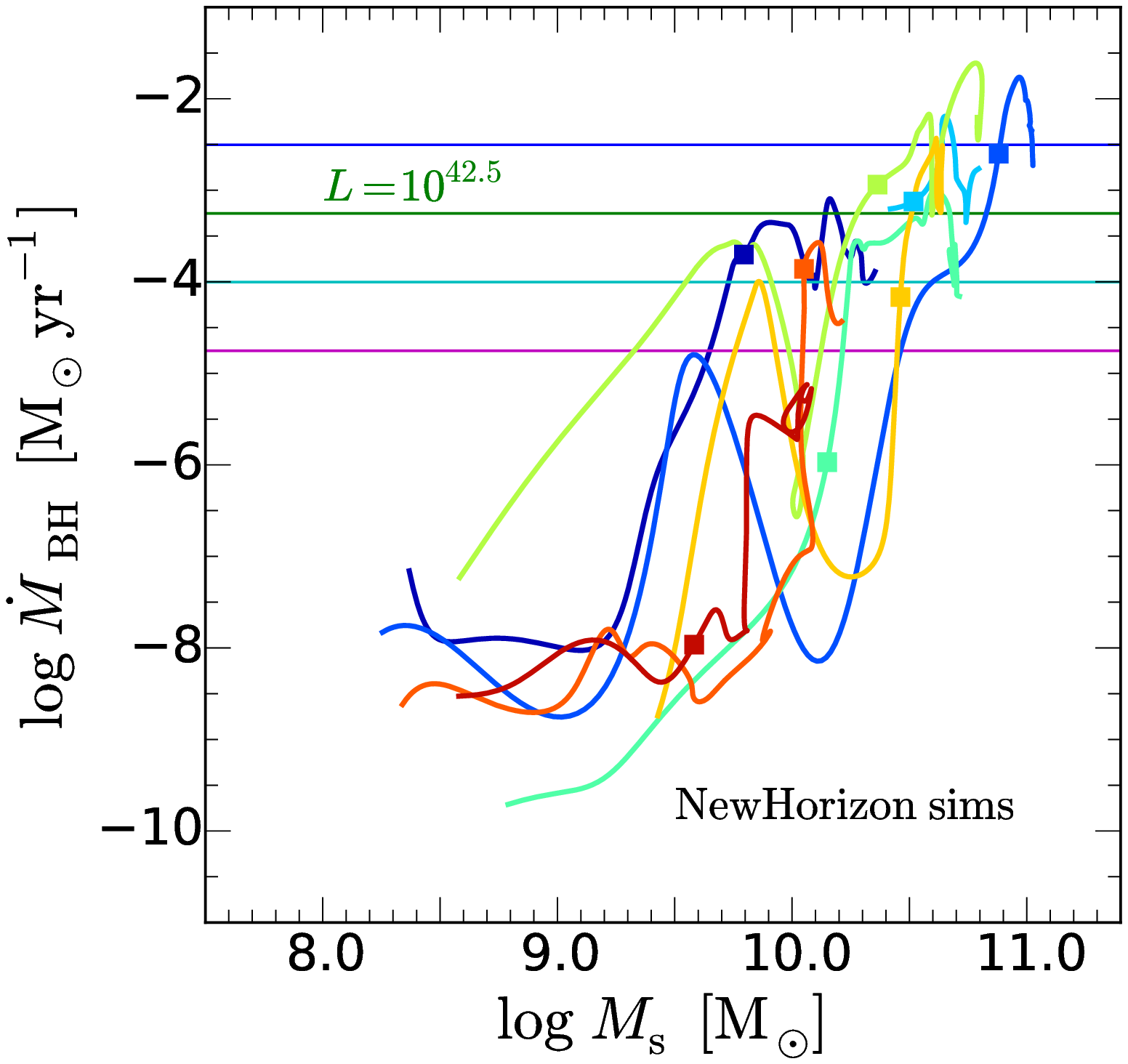}
\includegraphics{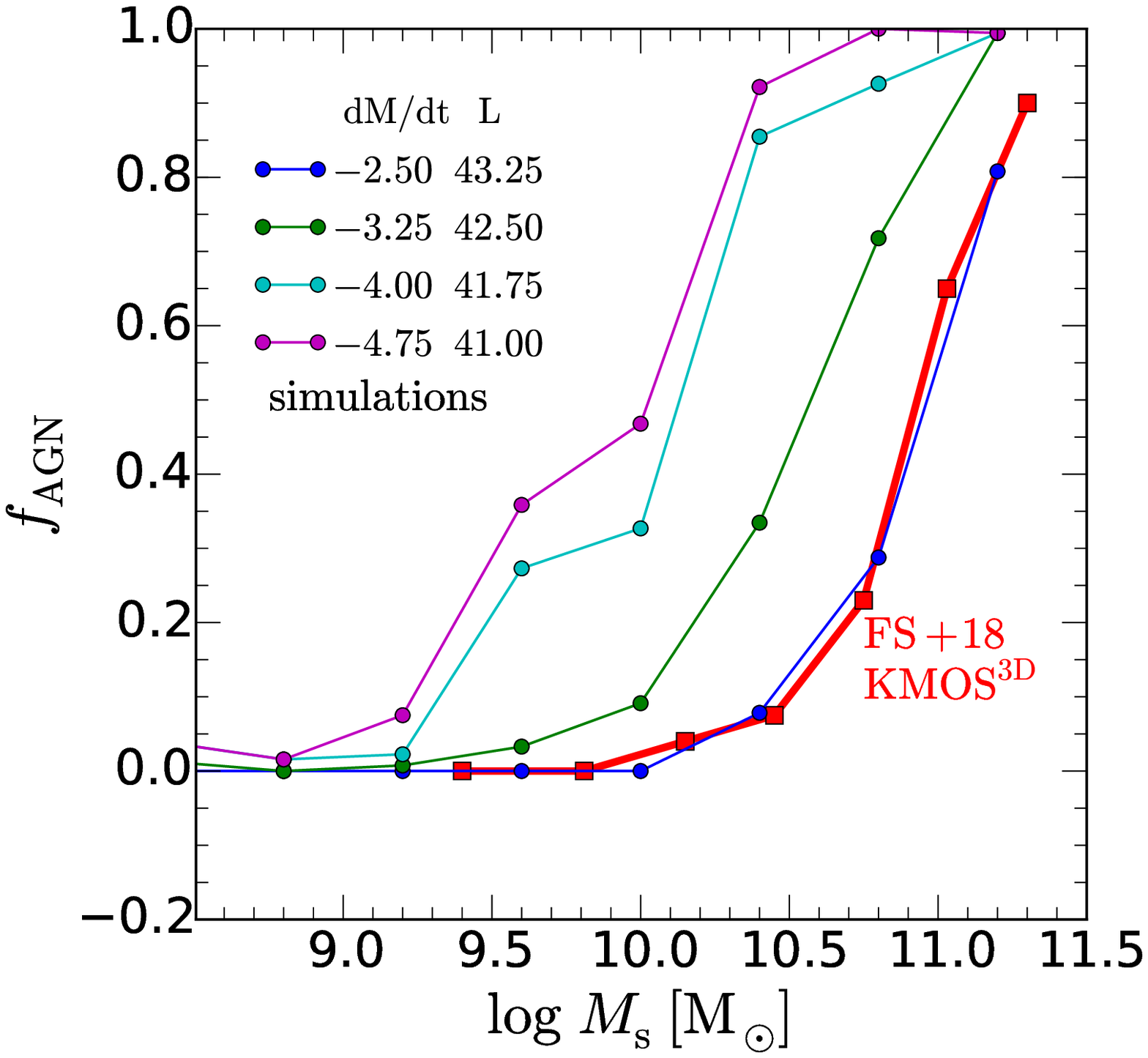}
\caption{
{\bf Left:}
Accretion rate onto the black hole versus $\Ms$ in New-Horizon simulations
\citep{lapiner19},
showing L-shape tracks with a turnover near the golden mass,
similar to \fig{Mbh} (right).
{\bf Right:}
When AGN are selected above a threshold luminosity (lines in the left panel
with $L=0.1 \dot{M} c^2$),
the fraction of galaxies that host AGN is plotted versus $\Ms$.
A threshold in the range $10^{42.5-43.25} \ergs$ (green-blue)
seems to match the observed AGN fractions \citep[red curve,][]{forster18b}.
}
\label{fig:Mdotbh}
\end{figure*}

\smallskip
\Fig{kocevski}, similar to \citet{kocevski17},
shows the fraction of CANDELS-survey galaxies that host AGN in
the plane of SFR versus compactness, 
as measured by U-V color and stellar surface brightness within $1\kpc$,
$\Sigma_{1}$, respectively.
This plane is similar to the plane shown in \fig{L_shape}, where the L-shape
evolution tracks are detected, except that the SFR is growing downwards.
One can indeed see a high AGN fraction in the quadrant of high SFR and high
compactness representing the blue nugget phase of evolution, as predicted.
Similar evidence comes from obscured AGN that tend to reside in compact
star-forming galaxies \citep{chang17}.

\smallskip
In order to reproduce the fraction of galaxies that host AGN as a function of
their stellar mass based on the New-Horizon simulations, 
we plot in the left panel of \fig{Mdotbh} the accretion 
rates onto the black hole as a function of $\Ms$ 
for the eight massive galaxies.
By assuming an AGN luminosity of $L\!=\!0.1\, d\Mbh/dt\, c^2$,
one can apply a luminosity threshold to select AGN for a sample.
The resultant AGN fraction, for each given threshold, is shown in the right
panel of \fig{Mdotbh}, in comparison to the fraction observed by
\citet{forster18b} using an effective luminosity threshold of $10^{42.5}\ergs$
(red curve).
The predicted curves all resemble in their general shape the observed curve, 
though the turnover mass is shifted as a function of the luminosity threshold.
When assuming the same threshold as in the observed sample (green),
the turnover mass is apparently underestimated by 0.4dex, while a threshold of
$10^{43.25}\ergs$ (blue) provides a perfect fit.
It is not clear whether this offset reflects an uncertainty in the luminosity
threshold applied in the observations and its correspondence to accretion rate
or a real underestimate of the golden mass in the simulations. 
We should note that while the existence of a golden mass is robust, rooted 
in the physical phenomena discussed above, the exact value of the golden mass 
in each simulation may be subject to the way the subgrid physics of supernova 
feedback and black-hole growth are implemented.

\smallskip
\citet{forster18b} also provide in their figure 9
an indication for an anti-correlation between
the AGN fraction and the stellar effective radius for galaxies of
$\Ms\!\sim\! 10^{10.00-10.75}\msun$, consistent with the predicted
compaction-driven AGN activity at and above the golden mass.
Their figure 3 indicates a correlation of AGN fraction with $\Delta$MS, the
deviation from the ridge of the Main Sequence of star-forming galaxies at
a given stellar mass, consistent with the onset of AGN activity at the
blue-nugget phase, where the galaxy is above the Main-Sequence ridge
\citep{tacchella16_ms}, similar to what is seen in \fig{kocevski}.

%%%%%%%%%%%%%%%%%%%%%%%%%%%%%%%%%%%%%%% 7
\section{Conclusion}
\label{sec:conc}

% golden scale obs
Observations reveal a golden mass scale for efficient star formation in
galaxies within dark-matter halos, at $\Mv \!\sim\! 10^{12}\msun$.
We point out that this translates to the golden time for star formation at
$z\!\sim\! 2$ because the (Press-Schechter) mass of typical forming halos at 
that epoch is comparable to the golden mass.
The apparent threshold for luminous AGN at a similar mass, while there is no
hint in black-hole physics for a characteristic mass of this sort,
indicates that it is imprinted on the central black hole by processes
associated with galaxy evolution.

\smallskip % confinement by two physical processes
Two physical processes confine the golden mass.  On the low-mass side,
energetic considerations imply that supernova feedback is effective in
suppressing star formation for $M \!<\! \Mcrit$ \citep[e.g.][]{ds86}.
On the high-mass side,
an analysis of virial shock stability reveals that
the halo CGM is heated to the virial temperature for $M \!>\! \Mcrit$ 
\citep[e.g.][]{db06}.
The resemblance of the critical masses associated with these two different
processes (\fig{scale_z}) confines efficient galaxy formation to a
peak about the golden mass.
% coincidence? 
One may wonder whether this similarity between the two critical masses is a
coincidence or a built in match. 
On one hand, the analyses in the two cases are related in the sense that the 
preferred scale arises from comparing a radiative cooling time to the relevant
dynamical time. On the other hand, these timescales refer to different 
environments on different scales. The conditions for cooling (e.g., gas 
density and metallicity) and the relevant dynamical times 
(in star-forming clouds versus at inflow into the halo) are very different.

\smallskip % compaction at the golden mass 
Cosmological simulations \citep{zolotov15} and observations
\citep{barro13,barro17} 
reveal that most galaxies undergo wet-compaction events throughout their
histories, events that induce major transitions in the galaxy properties.
In particular, the compactions trigger quenching of star formation by central
gas depletion, to star formation and outflows.
The compaction processes are due to drastic angular-momentum losses, 
about 40\% caused by mergers and the rest by counter-rotating streams,
recycling fountains and other mechanisms.
The major deep compaction events to blue nuggets, those that trigger a decisive
long-term quenching process and a transition from central dark-matter to baryon
dominance,
tend to occur near the golden mass, at all redshifts.
This is seen in simulations \citep{zolotov15,tomassetti16} and
in machine-learning-aided comparisons to observations \citep{huertas18}.
We argue that this preferred mass scale for major compactions is due to the 
same two physical processes of supernova feedback and hot CGM, which tend
to suppress compaction attempts at lower and higher masses.

\smallskip % threshold for BH growth
We argue that the characteristic threshold mass for rapid black-hole growth 
and AGN reflects the golden mass of galaxy formation.
Black-hole growth is suppressed by supernova feedback in star-forming galaxies
of mass below the golden mass, pre-compaction.
Black holes can grow above the golden mass as the gas is confined to the center
by the halo potential well and hot CGM.
In between the supernova and CGM phases,  
the onset of rapid black growth is driven by the major compaction event,
near the golden mass.

\smallskip % quenching -- trigger and maintenance
The quenching of star formation at $M\!>\!\Mcrit$ is thus a multi-stage process.
It is triggered by a major compaction event, via central gas depletion due to
star formation and the associated outflows.
The quenching is then maintained by the hot CGM that suppresses the cold gas
supply in $M\!>\!\Mcrit$ halos, especially at $z\!<\!2$ when cold streams hardly
penetrate the hot CGM. 
Finally, AGN feedback, triggered by the compaction-driven black-hole growth,
helps keeping the CGM hot and maintaining long-term quenching.  
Note that in this advocated scenario,
contrary to a common belief,
AGN feedback is not responsible for the onset of quenching. Instead, they
both result from the same wet compaction event once the galaxy is near the
golden mass. AGN feedback kicks in later on as a source of quenching
maintenance.
In other words, AGN feedback is not responsible for the golden mass and epoch
for galaxy formation, but it helps emphasizing it by further heating the gas
post compaction while in the hot CGM zone and after departing the SN zone.

\smallskip % interplay galaxy and center
It may be interesting to consider the interplay between the galaxy on large
scales and its central region during the different phases.
In the star-forming supernova regime below the golden mass, cold flows that
originate
from outside the halo induce star formation that is regulated by supernova
feedback, which suppresses the central black-hole growth.
Near the golden mass, major compaction events that are induced externally by
mergers and counter-rotating streams form compact star-forming blue 
nuggets, which trigger central quenching inside out. The compaction outside-in
causes a rapid central black-hole growth.
Above the golden mass, the hot CGM maintains the quenching by suppressing
the cold gas supply from the halo to the center.  
The halo deep potential well and hot CGM lock the supernova ejecta at
the galaxy center and allow a continuous black-hole growth. 
The resultant central AGN helps the quenching on larger scales by keeping the
CGM hot.
There is thus an strong two-way cross-talk between the galaxy at large, its
dark matter halo and the cosmic web, and the galactic center with its
supermassive black hole.

%%%%%%%%%%%%%%%%%%%%%%%%%%%%%%%

%\begin{acknowledgements}
%Thanks
%\end{acknowledgements}

\bibliographystyle{aa} % style aa.bst
\bibliography{bn} % your references Yourfile.bib

\end{document}